\documentclass{acmsmall}

\usepackage[utf8]{inputenc}
\usepackage[T1]{fontenc}
\usepackage{bibentry}
\usepackage{tabularx}
\usepackage[table]{xcolor}
\usepackage{cite}
\usepackage{comment}
\usepackage[colorlinks,linkcolor=blue,citecolor=blue,urlcolor=blue]{hyperref}

\title{The Multiple Facets of Software Diversity: Recent Developments in Year 2000 and Beyond}
\author{Benoit Baudry, Martin Monperrus\\
INRIA and University of Lille\\
benoit.baudry@inria.fr, martin.monperrus@univ-lille1.fr\\~\\
}

  \begin{abstract}
  Early experiments with software diversity in the mid 1970's investigated N-version programming and recovery blocks to increase the reliability of embedded systems. Four decades later, the literature about software diversity has expanded in multiple directions: goals (fault-tolerance, security, software engineering); means (managed or automated diversity) and analytical studies (quantification of diversity and its impact). 
  Our paper contributes to the field of software diversity as the first paper that adopts an inclusive vision of the area, with an emphasis on the most recent advances in the field. 
    This survey includes classical work about design and data diversity for fault tolerance, as well as the cybersecurity literature that investigates randomization at different system levels. 
It broadens this standard scope of diversity, to include the study and exploitation of natural diversity and  the management of diverse software products.    
    Our survey includes the most recent works, with an emphasis from 2000 to present. 
    The targeted audience is researchers and practitioners in one of the surveyed fields, who miss the big picture of software diversity. 
    Assembling the multiple facets of this fascinating topic sheds a new light on the field.
\end{abstract}

\begin{document}

\maketitle

\section{Introduction}

In nature, diversity refers to the fact that many species coexist (among many other definitions). In society, it sometimes refers to the idea of gathering people coming from different cultures and background.
In all these domains, diversity (a fact) is considered essential for the emergence of resilience, stability or novelty (a property) \cite{mccann00}. 
In software, we take the problem upside-down. We want properties, e.g. resilience, for which diversity may be the key. The main research question is thus formulated as: how to create, maintain, exploit -- i.e. engineer -- diversity in software? 

For instance, early experiments with software diversity in the mid 1970's (e.g. recovery blocks \cite{randell75}) advocate design and implementation diversity as a means for tolerating faults. Indeed, similarly to natural systems, software systems including diverse functions and elements are able to cope with many kinds of unanticipatable problems and failures.
Currently, the concept of software diversity appears as a rich and polymorphic notion, with multiple applications. Yet, the exploration of this concept is very fragmented over different communities, who do not necessarily know each other. 

We aim at putting together the many pieces of the puzzle of software diversity. 
Previous surveys on classical work about diversity for fault-tolerance \cite{deswarte98} or for security \cite{just04} provide important milestones in this direction. Yet, their scope is very focused on a single type of software diversity and they do not include the most recent works in the area.
Our paper contributes to the field of software diversity, as the first paper that adopts an inclusive vision of the area, with an emphasis on the most recent advances in the field. 

\paragraph{Scope}
This survey includes classical work about design and data diversity for fault tolerance, as well as the cybersecurity literature that investigates randomization at different system levels. 
Beyond that, we broaden this standard scope of diversity, to include work about the study and exploitation of natural diversity and about the management of diverse software products in software architecture. 
Since the main barriers between communities are words, we had to cross terminological chasms several times: diversity, randomization, poly- and meta-morphism, to only cite a few that are intrinsically related.
This inclusive definition allows us to draw a more complete landscape of software diversity than previous surveys \cite{Knight2011,Schaefer2012,just04,deswarte98}, which we discuss in section \ref{sec:other-surveys}.
For the first time, this survey  gathers under the same umbrella works that are often considered very different, while they share a similar underlying concept: software diversity. 

\paragraph{Novelty}
The field of software diversity has been very active in the 70's and 80's for fault-tolerance purposes. There has been a revival in the late 90's, early 2000's, this time with automatic diversity for security. Both periods have been covered by previous surveys \cite{deswarte98,just04}. 
The last decade's research on software diversity has also been extremely rich and dynamic. Yet, this activity is only partially covered in recent surveys by Schaeffer et al. \cite{Schaefer2012}, Knight \cite{Knight2011} and Larsen et al. \cite{Larsen14}, which have specific focuses. 
Our survey includes the most recent works in all areas of software diversity, with an emphasis from 2000 to present. 
 
\paragraph{Audience}
The targeted audience of this paper is researchers and practitioners in one of the surveyed fields, who miss the big picture of software diversity. 
Our intention is to let them know and understand the related approaches, so far unknown to them because of the community boundaries. 
We believe that this shared awareness and understanding, with different technical backgrounds, will be the key enabling factor for the development of integrated and multi-tier software diversification techniques \cite{allier14}. This will contribute to the construction of future resilient and secure software systems.

\begin{figure}[h]
	\begin{center}
		\includegraphics[width=\columnwidth]{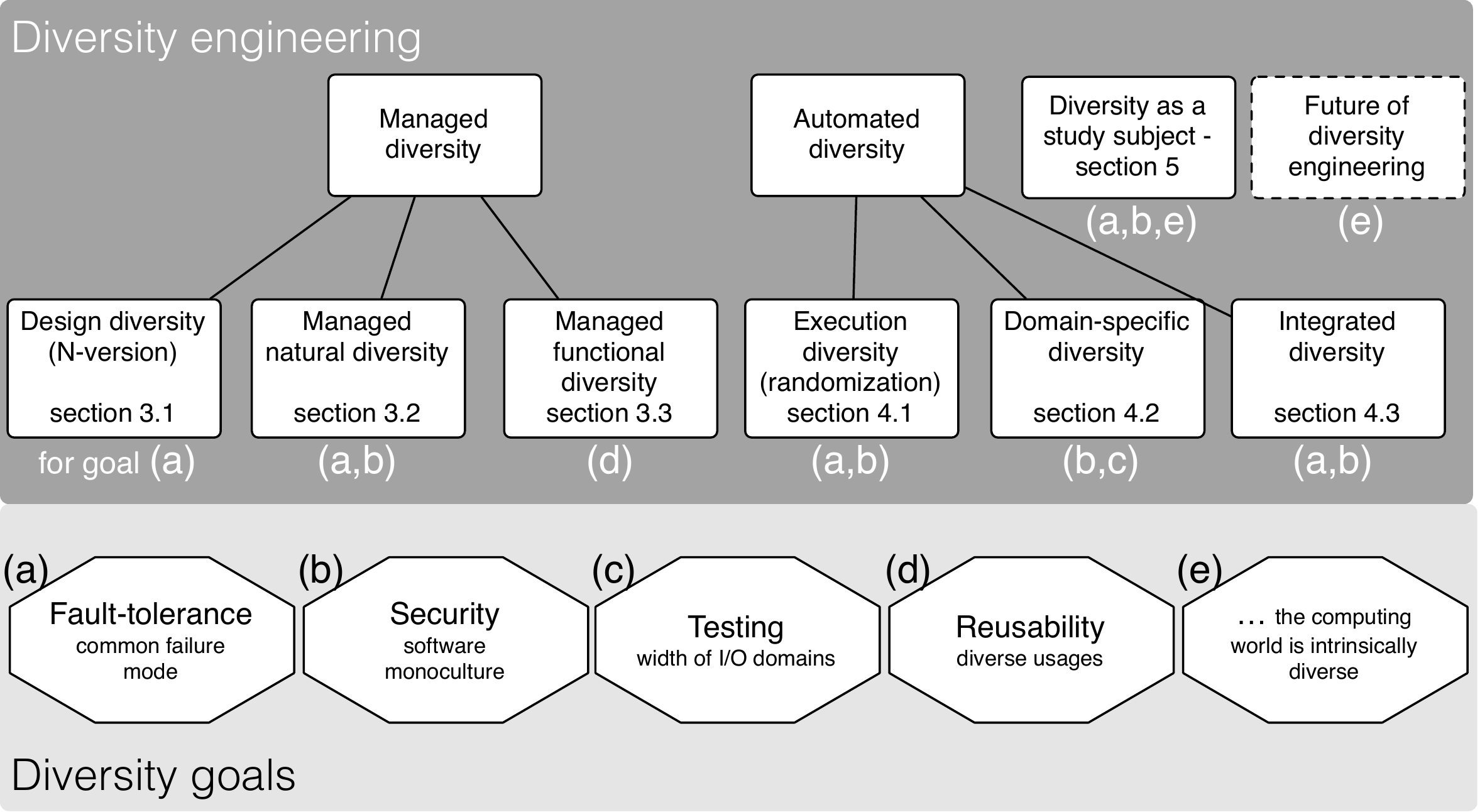}
		\caption{The diverse dimensions of software diversity}
        \label{fig:global-map}
	\end{center}
\end{figure}

\paragraph{Structure}
Given the breadth of this work's scope, there is no single decomposition criterion to structure our paper. Software diversity has multiple facets: the goal of diversity, the diversification techniques, the scale of diversity, the application domain, when it is applied~\ldots 
This diversity of software diversity is reflected in table \ref{tab:diversities}. 
As shown in Figure \ref{fig:global-map}, we decide to organize this survey mainly along two oppositions. 
First, we differentiate engineering work that aims at exploiting diversity (Sections \ref{sec:managed} and \ref{sec:automated-diversity}) from papers that are more observational in nature, where software diversity is a study subject (Section \ref{sec:natural-study}).
Then, we split the engineering papers on 
\emph{managed diversity} approaches, that aim at manually controlling software diversity
(section \ref{sec:managed}); 
and the papers describing  \emph{automated diversity} techniques (section \ref{sec:automated-diversity}).
This structuring supports our main goal of bridging different research communities and enables us to discuss, in the same section, papers coming from very different fields. 
The paper can be read linearly. However, each section is meant to be self-contained and there is a diversity of reading pathways.
We invite the reader to use Figure \ref{fig:global-map} for choosing her own one.

\begin{table}
\tbl{The diversity of software diversity (not exhaustive overview). Over time and over research communities, many kinds of software diversity have been proposed or studied.}{
\begin{tabularx}{\textwidth}{p{4cm}|X}
Software diversity for \ldots & Fault tolerance \cite{randell75,avizienis84}, security \cite{Forrest:1997,cox06}, reusability \cite{pohl2005software}, software testing \cite{chen2010adaptive}, performance \cite{Sidiroglou-Douskos2011}, bypassing antivirus
software \cite{BorelloFM10} \ldots\\
\hline
Software diversity at the scale of \ldots & Networks  \cite{donnell04}, operating systems \cite{koopman99}, components \cite{gashi04}, data structures \cite{ammann88}, statements \cite{Schulte2013mutrob}, \ldots\\
\hline
Software diversity as \ldots & a natural phenomenon \cite{mendez13}, a goal \cite{cohen93}, a means \cite{collberg12}, a research object \cite{knight86} \ldots\\
\hline

Software diversity in \ldots & market products \cite{han09}, operating systems \cite{koopman99}, developer expertise \cite{posnett13}, \ldots\\
\hline

Software diversity when \ldots & the specifications are written \cite{Yoo2002111}, the code is developed \cite{avizienis84}, the application is deployed \cite{franz10}, executed \cite{ammann88} \ldots\\

\end{tabularx}}
\label{tab:diversities}
\end{table}

\section{Survey Process} 

To prepare this survey, we first analyzed the existing surveys on the topic (see Section \ref{sec:other-surveys}). None of them covers the material we cover. Second, we set up and conducted a systematic process described in \ref{sec:process} 

\subsection{Other Surveys on Software Diversity} 
\label{sec:other-surveys}

The oldest survey we found is by Deswarte et al. in 1998 \cite{deswarte98}.
It clearly shows that software diversity has different scales:
from the level of human users or operators to the level of hardware and execution.
Our survey exactly goes along this line of exploring the diversity of diversities.
In addition to classical and 90ies' software diversity, our survey discusses the rich work that has been done around software diversity during the last fifteen years: instruction-set randomization, adaptive random testing, and many others.

In 2001, Littlewood et al. \cite{littlewood01} focus on design diversity (N-version programming). They review in particular their own work on the probabilistic reasoning that can be made on N-version systems. To this extent, as the abstract puts it, the survey is more a tutorial on design diversity than a broad perspective on software diversity.

The goal of Just et al.'s review paper  \cite{just04} is to list the techniques of synthetic diversity that can improve software survivability.
``Synthetic diversity'' is equivalent, in our views, to ``artificial automated diversity''.
In our paper, we consider other goals than only security (such as quality of service, see section \ref{sec:unsound}), and consider other diversity engineering techniques (e.g., managed software diversity, see \ref{sec:managed}).

John Knight  published a survey in 2011 \cite{Knight2011}.
He discusses four kinds of diversity:
classical design diversity (N-version and recovery block),
data diversity (a research direction he has both invented and lead),
artificial diversity (in the sense of instruction-set randomization for security and the like),
and N-variant systems (compared to N-version, N-variant diversity uses artificial and automated diversity).
In addition, he introduces the concept of ``temporal diversity'' as a diversity over time, for instance by regularly changing the key for instruction-set randomization.
We agree on all points that Knight considers as software diversity. However, we have a broader definition of software diversity: we discuss more kinds of managed software diversity (such as software product lines, see \ref{sec:spl}), 
more kinds of artificial diversity (such as runtime diversity, see section \ref{sec:dyn-randomization}), 
and papers for which diversity is the main study subject (see Section \ref{sec:study-subject}).

Schaefer and colleagues co-authored in 2012 ``Software diversity: state of the art and perspectives'' \cite{Schaefer2012}.
Despite what the title suggests, this paper surveys only one kind of software diversity: software product lines.
As we will discuss later, the techniques of software product lines  enable one to manage a set of related features to build diverse products in a specific domain. We refer to this kind of diversity as ``managed software diversity''. 
In our paper, not only do we describe other kinds of managed software diversity such as design diversity, but we also discuss artificial diversity and natural diversity as well.

Larsen et al. \cite{Larsen14} recently authored a survey about automated software diversity for security and privacy. They discuss  the different threat models that can be addressed via diversification. Then, they classify the surveyed approaches according to the nature of the object to be diversified and the temporal dimension of the diversification process. They conclude with an insightful discussion about compiler-based vs. binary rewriting diversity synthesis.

\subsection{Systematic Process} 
\label{sec:process}
We followed a systematic process to select the papers discussed in this paper. We started with 30 papers that we knew and are written by the most remarkable authors: Avizienis, Randell, Forrest, Cohen, Knight and Levenson, Schaeffer, etc.. They appear in top publications of these fields (ACM TISSEC, IEEE TSE, IEEE S\&P, CCS, ICSE, PLDI, DSN, etc.) and are generally considered as seminal work in each area. Then, we increased this set through a systematic keyword-based search using Google Scholar, IEEE Xplore and ACM DL. This set went through a second expansion phase when we followed the citation graph of the selected papers. This provided us with a set of more than 300 papers. Then, we filtered out papers. First, we discarded the redundant papers that discuss a similar problem or solution (e.g., we selected only a few papers about product lines or about multi-version execution). Second, we filtered out the papers that had no impact on the literature (that appear in unknown conferences or that had less than 5 citations after 20 years). Since our survey focuses on recent developments in the field of software diversity, we took a special care to keep the most significant recent works (up to papers that appeared in 2014).

\section{Managed Software Diversity}
\label{sec:managed}

``Managed  software diversity'' relates to technical approaches aiming at encouraging or controlling software diversity.
This kind of diversity is principally embodied in the work on multi-version software (early structuring of diversity), open software architecture (encouraging diversity) and software product lines (controlling diversity).

\subsection{Design Diversity (N-Version)}
\label{sec:design}

Since the late 1970's many different authors have devised engineering methods for software diversification to cope with accidental and deliberate faults. Here, an accidental fault is any form of bug, \textit{i.e.}, an internal problem unintentionally introduced by a developer of the execution environment. N-version programming  \cite{avizienis85} and recovery blocks \cite{randell75} were the two initial proposals to introduce diversity in computation to limit the impact of bugs. Those techniques are traditionally called ``design diversity'' techniques. 
 
N-version design is defined as ``the independent generation of $N\geq2$ functionally
equivalent programs from the same initial specification''  \cite{avizienis84,avizienis85}. This consists in providing N development teams with the same requirements.
Those teams then develop N independent versions, using different technologies, processes, verification techniques, etc. The N versions are then run in parallel and a voting mechanism is executed on the N results.
The increased diversity in design, programming languages and humans is meant to reduce the number of faults by emergence of the best behavior, the emergence resulting from the vote on the output value. 

Since the initial definition of the N-version paradigm, it has been refined along different dimensions: the process, the product and the environment necessary for N-version development \cite{avizienis95}. For example Kelly \cite{kelly91} distinguishes between random diversity (let independent teams develop their version) from enforced diversity in which there is an explicit effort to design diverse algorithms or data structures. More recently, Avizienis proposed to adapt the concept to software survivability \cite{avizienis00}.

Recovery blocks were developed at the same time as N-version design, and proposed a way of structuring the code, using diverse alternative software solutions, for fault tolerance \cite{randell75}. The idea is to have recovery blocks in the program, \textit{i.e.}, blocks equipped with error detection mechanisms and one or more spares that are executed in case of errors. These spares are diverse variant implementations of the function. 

In the latest work about N-version development, both N-version design and recovery blocks were included in the same global framework \cite{avizienis95}. This framework has then been used in multiple domains, including the design of multiple versions of firewalls \cite{liu2008}. 
While the essential conceptual elements of design diversity have remained stable over time, most subsequent works have focused on experimenting and quantifying the effects of this approach on fault tolerance. The work related to the analysis of N-version programming is synthesized in section \ref{sec:study-n-version}.

\subsection{Managed Natural Software Diversity}
\label{sec:managed-natural}
We call ``natural diversity'', the existence of different software solutions that provide similar functionalities and which spontaneously emerge from software development processes.
There exists several forms of natural software diversity. For example, the programs that can be customized through several parameters, embed a natural mechanism for diversification (two instances of the same program, tuned with different parameters can have different behaviors in terms of performance). Software market and competition are also strong vectors that drive the natural emergence for software diversity. For example, the gigantic business opportunities offered by the world wide web has driven the emergence of many competing web browsers. Web browsers are diverse in their implementation, in their performance, in some of their plugins, yet they are functionally very similar and can be used for one another in most cases. Other examples of such market software diversity include operating systems, firewalls, database management systems, virtual machines, routers, middleware, application servers, etc. In this section we present a set of works which exploit this natural diversity for different purposes. We will come back to natural diversity later in Section \ref{sec:natural-study}, for discussing authors who study natural diversity with no engineering goals at all.

Hiltunen et al. \cite{Hiltunen00} propose the Cactus mechanism for survivability, i.e., a mechanism that monitors and controls a running application in order to tolerate unpredictable events such as bugs or attacks. The Cactus approach relies on fine grain customization of the different components in the application, as well as runtime adaptation, to achieve survivability. They discuss how they can switch between different security and fault-tolerance solutions through customization and they also discuss how this natural way of changing a system supports the emergence of natural diversity and thus increases resilience.

Caballero et al.  \cite{caballero08} exploit the existing diversity in router technology to design a network topology that has a diverse routing infrastructure. Their work introduces a novel metric to quantify the robustness of a network. Then, they use it to compare the robustness of different, more or less diverse, routing infrastructure. They explore the impact of different levels of diversity, by converting the problem into a graph coloring problem. They show that a small amount of router technology and well designed topology actually increases the global robustness of the infrastructure.

Totel et al. \cite{totel06} propose to design an intrusion detection mechanism by design diversity, leveraging the natural diversity of components-off-the-shelf (COTS). They exploit the fact that COTS for database management  and web servers have very few common mode failures \cite{wang03,gashi04} and are thus very good candidates for N-version design based on natural diversity. The authors deploy an architecture with three diverse servers running on three different operating systems and feed it with the requests sent on their campus web page in the last month (800000 requests, out of which around 1\% can be harmful). The results show that the COTS-based IDS only raises a small number of false positives.  Along the same line, Garcia et al. \cite{garcia2014analysis} conducted a study on the impact of operating system diversity w.r.t. to security bugs of the NIST National Vulnerability Database (NVD). Their results show that diversity indeed contribute to building intrusion-tolerant systems.

Oberheide et al. \cite{oberheide08} exploit the diversity of antivirus and malware systems to propose what is called ``N-version protection''. It is based on multiple and diverse detection engines running in parallel. Their prototype system intercepts suspicious files on a host machine and send them in the cloud to check for viruses and malware against diverse antivirus systems. They evaluate their system over 7220 malware and show that it is able to detect 98\% of the malware. It provides better results than a single antivirus in 35\% of the cases. The idea has been further explored by Bishop et al. \cite{bishop2011diversity}, who explored the deep characteristics of the dataset of known malware to reduce global vulnerability.

O'Donnell and Sethu \cite{donnell04} leverage the diversity of software packages in operating systems and investigates several algorithms to increase the global diversity in a network of machines. They model the diversification of distributed machines as a graph coloring problem and compare different algorithms according to their ability of setting a network that is tolerant to attacks. The experiments are based on a simulation, which uses the topology from email traffic at the authors' institution. They show that the introduction of diversity at multiple levels provides the best defense.

Carzaniga et al. \cite{carzaniga10}  find multiple different sequences of method calls in Javascript code, which happen to have the same behavior. They harness this redundancy to setup a runtime recovery mechanism for web applications.  

Gorbenko et al.  \cite{gorbenko11} propose an intrusion avoidance architecture based on multi-level software diversity and dynamic software reconfiguration in IaaS cloud layers. The approach leverages the natural diversity of off-the-shelf components that are found in the cloud (operating system, web server, database management system and application server), in combination with dynamic reconfiguration strategies. The authors illustrate the approach with an experiment over several weeks, during which they switch between 4 diverse operating systems that have different open vulnerabilities. They discuss how this mechanism reduces exposure to vulnerabilities.

\subsection{Managed Functional Diversity}
In software, it is known that many functions are the same yet different. 
For instance, passing a message to a distant machine or writing to a local file is conceptually the same: writing data to a location. However, the different implementations (say for network or for file input/output)  of this abstract function are radically different.
One responsibility of software abstractions is to capture this conceptual identity and to abstract over the diversity of implementation details.
For instance, Unix is well known because of Unix' concept of file captures all input/output operations, whether on the network, on a physical file on disk or on the memory of a kernel module.
We refer to this facet of abstraction as managing the functional diversity.

Many software abstractions have the clear goal of managing functional diversity.
In the following, we will review classical object-oriented software, software product lines and plugin-based architecture.

\subsubsection{Class Diversity}
The object-oriented software paradigm is a rich paradigm with implications on understandability, reuse, etc. 
There is one point in this paradigm really related to managing the diversity: polymorphism.

Polymorphism is the mechanism enabling us to have code that calls other pieces of code in a non predefined manner. The late binding between functions enables an object to call a diverse set of functions and even to call code that will be written in the future. To this extent, polymorphism is the key mechanism enabling to manage  the function diversity (as embodied in classes). 
In other words, polymorphism (with abstract methods, interfaces or other fancy object-oriented constructs) supports the construction of a program architecture that is ready for handling diversity. 

As Bertrand Meyer~\cite{meyer1988object} puts it:
\begin{quote}
\emph{``We are at the heart of the object-oriented method's contribution to reusability: offering not just frozen components (such as found in subroutine libraries), but flexible solutions that provide the basic schemes and can be adapted to suit the needs of many diverse applications.''}
\end{quote}

\subsubsection{Software product lines}
\label{sec:spl}

The techniques around software product lines can be considered as means of controlling a diversity of software solutions capable of handling a diversity of requirements (user requirements or environmental constraints) \cite{pohl2005software,clements02}.
Software product line engineering  is about the development of ``\emph{a diversity of software products and software-intensive systems at lower costs, in shorter time, and with higher quality}'' \cite{pohl2005software}. This consists in building an explicit variability model, which captures all commonalities and variation points in requirements and software solutions. In other words, the variability model is an explicit definition of the space of diverse solutions that can be engineered in a particular domain. This model is usually expressed as a form of feature model \cite{kang90}. 

In the context of software product lines, the main challenge for software diversity management consists in providing systematic ways to reuse existing parts of software systems in order to derive diverse solutions. 

We synthesize the main works in software product lines, for an exhaustive survey, we refer the reader to Schaefer et al.’s survey ``Software diversity: state of the art and perspectives'' \cite{Schaefer2012}. We start by looking at solutions that  handle diversity in design, then we summarize solutions for diversity in implementation. 

Software product lines mainly offer support for design diversity through architectural solutions \cite{clements02}. An essential challenge is to handle both the logical variability (the set of features that architects manipulate) and the variability of concrete assets (diversity of software pieces that can actually be composed to implement a particular product). Initial solutions are based on annotations to relate both views \cite{atkinson02}. Hendrikson et al. \cite{hendrickson07} propose a product line architecture modeling approach that unites the two, using change sets to cluster related architectural differences. Several approaches are founded on a compositional approach to derive products from architectural models. Ziadi et al. \cite{ziadi04} propose sound composition operations for UML 2.0 scenarii in order to automatically synthesize diverse statecharts inside a given product line, while Morin et al. \cite{morin08} compose software components to derive software configurations at runtime. Other approaches rely on an orthogonal variability model associated to  model transformations for product derivation, as is the case for the Common Variability Language \cite{haugen08} or the Orthogonal Variability Model \cite{pohl2005software}.
At the boundary between models and implementation, it is possible to capture the variants of a program with explicit design patterns, as suggested by J\'ez\'equel \cite{jezequel1998}.
At the source code level, there exist several mechanisms to manage a set of variants for a given program: delta-oriented programming \cite{schaefer2010} instantiates the concept of delta-modeling \cite{clarke2011} to specify a specific set of deltas for a program, as well as transformations that can systematically inject a set of selected deltas in a program to derive a variant; Figueiredo and colleagues have reported on  the usage of aspect-oriented programming to hanlde variants in a product line and discuss the postive and negative effects on design stability \cite{figueiredo2008}; preprocessing was one of the first language technology used to handle program variants and has been extensively analyzed, for example in the recent work by Liebig et al. \cite{liebig2010}.

\subsubsection{Diversity through Plugin- and Component- based Software Architecture}

Plugin-based software architectures offer means to design open software systems. Plugins are software units that encapsulate a given functionality as well as some information about its dependencies. As far as we know, Wijnstra \cite{wijnstra00} was one of the first authors to assess the suitability of plugins to handle the diversity of configurations and usages of a complex software system \cite{wijnstra00}. He proposed to use plugins, together with a component framework to design an extensible system for medical imaging. In this context, he needed to have a core set of functionalities to deploy a diversity of products that fit different requirements or different environments. 

More recently, very successful software projects such as Wordpress, Firefox or Eclipse have adopted plugin-based architectures. This allows them to be open, thus leveraging the efforts of large open source communities, while keeping a core set of functionalities across all versions. But most importantly, this architecture supports a true explosion of functional software diversity. For example, there are 25000 plugins available for Wordpress, which can be combined by users in billions of functionally diverse configurations, each of them fitting a specific purpose or need. This was somehow predicted by Ommering \cite{ommering02}, who used a plugin-based architecture in which connections between plugins handle design-time or run-time diversity.

\subsubsection{Discussion}
The main benefit of those software construction paradigms with respect to diversity is reusability: a large range of diverse products can be made with a smaller number of software “bricks”.  This is our motivation for considering software construction and design paradigms in our survey.

However, the overall effect of those paradigms is to reduce software design diversity for a given set of product functions. Indeed, those reuse-oriented paradigms create a tension between reusability and monoculture \cite{allier14}. Both relate to diversity (the second one in a dual manner). In practice, there is an engineering tradeoff between the increase of diversity due to the infinite number of possible combinations and the decrease of diversity due to massive reuse.

\subsection{Summary}
This section has focused on three areas of software engineering, which \textit{manage} software diversity. The first was about multi-version design, an approach to fault-tolerance that  aims at managing the manual development of diverse program versions. The second part was about managing and exploiting software diversity that naturally emerges in software markets or open source communities, in order to build fault or attack tolerant systems. The last part opened on a series of works dedicated to the management of functional diversity, in order to fulfill the various usages of a given system. These three parts refer to different research communities, yet, they all share a common approach: software diversity can be managed and harnessed in order to achieve specific software engineering objectives.

\section{Automated Software Diversity}
\label{sec:automated-diversity}

``Automated software diversity''  consists of techniques for artificially and automatically synthesizing diversity in software.
Instead of using the adjective automated, some authors call it ``synthetic diversity'' \cite{just04} or ``artificial'' diversity (e.g. \cite{SidiroglouLK06}). However, artificial literaly means \emph{``created or caused by people''}\footnote{Merriam-Webster, \url{http://www.merriam-webster.com/dictionary/artificial}}. To this extent, N-version programming also produces artificial diversity but, the diverse program variants are produced manually.
We prefer ``automated diversity'' which emphasizes the absence of human in the loop and is in clear opposition to managed software diversity.
Beyond those details, we actually equate those three terms: artificial, synthetic and automated diversity.

Automated software diversity is valuable in different contexts, for instance software security or fault tolerance. However, these different \emph{goals} are not the only dimension in which we can characterize the various approaches to automated software diversity.
First, the \emph{scale} dimension characterizes the fact that software systems are engineered at several scales:
from a set of interacting machines in a distributed system down to the optimization of a particular loop.
Research has produced techniques for automated software diversity along all those different scales.
Second, the \emph{genericity} dimension explores whether the diversification technique is domain-specific or not. Third, the \emph{integrated} dimension is about the assembly of multiple diversification techniques in a global approach. 

\subsection{Randomization}

The mainstream software paradigms are built on determinism. 
All layers of the software stack tend to be deterministic, from programming language constructs, to compilers, to middleware, up to application-level code.

However, it is known that randomization can  be useful, for instance to improve security \cite{bhatkar03}. 
A classical example of randomization is compiler based-randomization: a compiler may compile the same code with different memory layouts to decrease the risk of code injection.

What is the relation between randomization and diversity? A randomization technique creates, directly or indirectly, set of unique executions for the very same program. As mentioned by \cite{bhatkar03}, \emph{``the use of randomized program transformations [is] a way to introduce diversity into applications''}. The notion of ``diversity of execution'' is broad: it may mean diverse performances, diverse outputs, diverse memory locations, etc. 
We present an overview of diversifying randomization techniques in this survey.
For a more detailed survey about randomization, we refer the reader to surveys dedicated to that topic, in particular the one of Keromytis and Prevelakis \cite{keromytis2005survey}.

There are different kinds of diversifying randomization. 
 First, one can create different versions of the same program.
For instance, one can randomize the data structures  at the source  or at the binary level. 
We call this kind of randomization ``static''. 
Static randomization is discussed in Section \ref{sec:static-randomization}. 

Second, one can automatically integrate randomization points in the executable program. 
For instance, a malloc primitive (memory allocation) with random padding is a randomization point: each execution of malloc yields a different result. 
Contrary to static randomization, there is still one single version of the executable program but their executions are diverse. We call this kind of randomization ``dynamic randomization'' (also called runtime randomization \cite{xu2003transparent}) and discuss it in \ref{sec:dyn-randomization}. 

Third, some randomization techniques do not aim at providing a strict behavioral equivalence between the the original program and the randomized executions. They are are discussed in Section \ref{sec:unsound}.

Finally, as we will see later in Section \ref{integrated-diversity}, diversification techniques can be stacked. This also holds for randomization: one can stack static and dynamic randomization. In this case, there are diverse versions of the same program which embed randomization points that themselves produce different executions.

\subsubsection{Static Randomization}
\label{sec:static-randomization}

One of seminal papers on static randomization is by Forrest and colleagues \cite{Forrest:1997}, who highlight two families of randomization: 
randomly adding or deleting non-functional code and
reordering code.
Those transformations are also described by Cohen \cite{cohen93} in the context of operating system protection. 
Lin et al. \cite{Lin2009} randomize the data structure of C code. Following the line of thought of Forrest et al. \cite{Forrest:1997} they re-order fields of data structures (\texttt{struct} and \texttt{class} in C/C++ code) and insert garbage ones. 

The concept of instruction-set randomization has been invented in 2003 in two independent teams \cite{Kc03,barrantes:03}
It consists of creating a unique mapping between artificial CPU instructions and real ones. This mapping is encoded in a key which must be known at runtime to actually execute the program. Eventually, the instruction set of a machine can be considered as unique, and it is very hard for an attacker ignoring the key to inject executable code.
Instruction-set randomization can be done statically (a variant of the program using a generated instruction set is written somewhere) or dynamically (the artificial instruction set is synthesized at load time).
In both cases, instruction-set randomization indeed creates a diversity of execution which is the essence of the counter-measure against code injection.

In some execution environments (e.g. x86 CPUs), there exists a ``NOP'' instruction. It means ``no operation'' and it has been invented for the sake of optimization, in order to align instructions with respect to some alignment criteria (e.g. memory or cache). 
Merckx \cite{merckxsoftware}  and later Jackson \cite{jackson} have explored how to use NOP to statically diversify programs. The intuition is simple: by construction ``NOP'' does nothing and the insertion of any amount of it results in a semantically equivalent program. However, it breaks the predictability of program execution and to this extent mitigates certain exploits.  

Obfuscation is a classical application domain of static randomization.
Code obfuscation consists of modifying software for the sake of hindering reverse engineering and code tampering. Its main goal is to protect intellectual property and business secrets.
A basic obfuscation technique simply transforms a program $P$ in a program $P'$ which is distributed. 
However, since obfuscation is automated, it is often possible to generate several different obfuscated versions of the same program (as proposed by Collberg et al. \cite{collberg12} for example). To this extent, code obfuscation is one kind of software diversification, with one specific criterion in mind.
For an overview on code obfuscation, we refer to the now classical taxonomy by Collberg and colleagues \cite{collberg1997taxonomy}.  For an example of a concrete obfuscation engine for Java programs, we refer to \cite{collberg1998manufacturing} and its Figure 1. 
When obfuscation happens at runtime, it is a kind of execution diversity and we discuss it in \ref{sec:dyn-randomization}.

\subsubsection{Dynamic Randomization}
\label{sec:dyn-randomization}

Chew and Song \cite{Chew02mitigatingbuffer} target ``operating system randomization''. More specifically, they randomize the interface between the operating system and the user-land applications:
the system call numbers, the library entry points (memory addresses) and the stack placement. All those techniques are dynamic, done at runtime using load-time preprocessing and rewriting. 

Dynamic randomization can address different kinds of problems. In particular, it  mitigates a large range of memory error exploits. Bathkar et al. \cite{bhatkar03,bhatkar2005efficient} have proposed some of the seminal research  in this direction. Their approach is based on three kinds of randomization transformations: randomizing the base addresses of applications and libraries  memory regions, random permutation of the order of variables and routines, and the random introduction of random gaps between objects. 

Static randomization creates diverse version of the same program at compilation time, dynamic randomization creates diverse executions of the same program under the same input at runtime. What about just-in-time compilation randomization? This point has been studied by Homescu and colleagues at the University of California Irvine \cite{homescu2013profile}. Their approach neither creates diverse versions of the same program nor introduces randomization points: the randomization happens in the just-in-time compiler directly. 
Their randomization is based on two diversification techniques: insertion of NOP instructions and constant blinding.

In the techniques we have just discussed, the support for dynamic randomization is implemented within the execution environment. On the contrary, self-modifying programs embed their own randomization techniques \cite{mavrogiannopoulos2011taxonomy}. This is done for sake of security and is considered one of the strongest obfuscation mechanism \cite{mavrogiannopoulos2011taxonomy}. 

Ammann and Knight's ``data diversity'' \cite{ammann88} represents another family of randomization. The goal of data diversity is not security but fault tolerance. The technique aims at enabling the computation of a program in the presence of failures. The idea of data diversity is that, when a failure occurs, the input data is changed so that the new input does not result in a failure. The output based on this artificial input, through a inverse transformation, remains acceptable in the domain under consideration. 
To this extent, this technique dynamically diversifies the input data. 

The notion of ``environment diversity''  \cite{vaidyanathan2005comprehensive} refers to techniques that change the environment to overcome failures. For instance, changing the scheduler or its parameter is indeed a change in the environment. This is larger in scope than just changing some process data, such as standard randomization.

\subsubsection{Unsound Randomization}
\label{sec:unsound}
Traditional randomization techniques are meant to produce programs or executions that are semantically equivalent to the original program or execution. 
However, have explored the domain of ``unsound'' randomization techniques, either statically or dynamically.

Foster and Somayaji \cite{Foster2010} recombine binary object files of commodity applications. If an application is made of two binary files A and B, they show that is possible to run the application by artificially linking a version of A with a different yet close version of B. The technique enables them to tolerate bugs and even let new functions emerging but has no guarantee on the behavior of the recombination.

Schulte et al. \cite{schultesoftware} describe a property of software that has never been reported before. Software can be mutated and at the same time, it can preserve a certain level of correctness. Using an analogy from genomics, they call this property ``software mutational robustness''.  This property has a direct relation to diversification: one can mutate the code in order to get functionally equivalent variants of a program. Doing this in advance is called ``proactive diversity''. The authors present a set of experiments that show that this proactive diversity is able to fix certain bugs.

In our previous work \cite{baudry14}, we experiment with different transformation strategies, on Java statements, to synthesize ``sosie'' programs. The sosies of a program P are variants of P, i.e., different source code, which pass the same test suite and that exhibit a form of computation diversity. In other words, our technique synthesizes large quantities of variants, which provide the same functionality as the original through a different control or data flow, reducing the predictability of the program's computation.

Another kind of runtime diversity emerges from the technique of loop perforation \cite{Sidiroglou-Douskos2011}.
In this paper, Sidiroglou et al. have shown that in some domains it is possible to skip the execution of loop iterations. For instance, in a video decoding algorithm (codec), skipping some loop iterations has an effect on some pixels or contours but does not further degrade or crash the software application. On the other hand, skipping loop iterations is key with respect to performance. In other words, there is a trade-off between the performance and accuracy. This trade-off can be set offline (e.g. by arbitrarily skipping one every two loops) or dynamically based on the current load of the machine.
In both cases, this kind of technique results in a semantic diversity of execution profiles, and consequently is deeply related to automated diversity. 

\subsubsection{Summary}
\label{sec:summary-randomization}

In this subsection, we have focused on techniques that automatically randomize some aspect of a program, thus producing a diversity of program versions. Diversity occurs in memory, in the operating system, in the bytecode or in the source code, but in all cases it happens with no human intervention, through random processes.

\subsection{Domain-specific Diversity}

The techniques we have presented so far are independent of any application domain. 
Yet, domain knowledge can be essential to devise efficient diversification techniques. This section illustrates such situations.

For instance, a common vulnerability of web applications is the possibility of injecting SQL code in order to access unauthorized data or corrupt existing one. Boyd et al. \cite{Boyd04} proposed a technique to diversify the SQL query themselves. By simply prefixing all SQL keywords with an execution specific token, they create an unpredictable language that is hardly attackable from the outside and diverse for each database. 

Feldt \cite{Feldt1998} exploited the structure of the genetic programming problem domain for the sake of diversification. He uses a genetic programming system to create a pool of diverse airplane arrestment controllers. He then shows that the failure modes of the synthesized programs are diverse, i.e. that the approach is effective for the generation of a kind of failure diversity.  

Oh et al. \cite{oh2002} presented a program transformation aiming at detecting a particular hardware fault (stuck-at faults in data paths of functional units). 
The transformation consists of multiplying all numerical computations by a constant $k$ in a semantics-preserving way. 
The authors show that this technique is effective with respect to their fault model. Obviously, it enables one to automatically obtain diverse implementations of the same program (for different values of $k$).

Computer viruses are programs whose main opponents are anti-virus systems. 
Inventors of computer viruses of course care about being reverse-engineered.
However, more importantly for them, the computer viruses must remain undetectable as long as possible. 
Diversification is one solution in this very specific domain: if the virus exists under many different forms, it is harder for anti-virus systems to detect them all. From the perspective of the virus itself, it is even better to constantly change itself. This kind of diversification is performed through so-called ``metamorphic engines'', where metamorphism refers to the concept of having different forms for the same identity. For a recent account on this kind of diversification we refer the reader to Borello and Mé \cite{BorelloFM10}.

In the domain of sensor networks, Alarifi and Du \cite{alarifi2006} propose an approach to diversifying sensor software in order to mitigate reverse engineering effort. Their approach diversifies both the data (e.g. the keys used to communicate between nodes) and the code. As a result, each node in a sensor network is very likely to be unique.

So far, we have discussed the diversification of software applications. 
Test cases are executable programs, but very specific ones. 
Although they are often written in general purpose programming languages,
their unique goal is to verify the correctness of an application. 
They do not provide services to users. 
Interestingly, this fundamental difference does not prevent diversity and diversification to be valuable in test cases as well. 
Adaptive random testing \cite{chen2010adaptive} is a random testing technique whose goal is generate input test data. 
It is adaptive in the sense that the generated test cases depend on the previously generated ones. The final goal is to evenly spread test cases throughout the input domain. 
To this extent, adaptive random testing aims at generating diverse test cases, and this is clear for the authors themselves, who subtitled their flagship paper: ``\emph{The art of test case diversity}''.
Feldt et al.'s VAT model is an example of adaptive random testing \cite{Feldt2008TestDiversity}.
They use an information distance for information theory to maximize the diversity of generated test cases.

\subsection{Integrated Diversity}
\label{integrated-diversity}

Integrated software diversity is about works that aim at automatically injecting different forms of diversity at the same time in the same program. 
In this line of thought, previous researchers have either emphasized the fact that the diversity is stacked (Section~\ref{sec:stacked-diversity}) or whether these different forms of diversity are managed with a specific diversity controller (Section~\ref{sec:controller-diversity}).

\subsubsection{Stacked Diversity}
\label{sec:stacked-diversity}

The different contributions discussed in this section all share the same intuition that each kind of artificial diversity has value in one perspective (a specific kind of attack or bug), and thus, integrating several forms of diversity should increase the global ability of the software system with respect to security or fault tolerance.

Wang et al. \cite{wang01} propose a multi-level program transformation that aims at introducing diversity at multiple levels in the control flow so as to provide in-depth obfuscation. This work on program transformation takes place in the context of a software architecture for survivable systems as proposed by Knight et al. \cite{knight00}. Wang et al's architecture relies on probing mechanisms that integrate two forms of diversity: in time (the probe algorithms are replaced regularly) and in space (there are different probing algorithms running on the different nodes of the distributed system). 

Bhatkar et al. \cite{bhatkar03} aim at developing a technique for address obfuscation in order to thwart code injection attacks. This obfuscation approach relies on the combination of several randomization transformations: randomize base addresses of memory regions to make the address of objects unpredictable; permute the order of variables in the stack; and introduce random gaps in the memory layout. Since all these transformations have a random component, they synthesize different outputs on different machines, thus increasing the diversity of attack surfaces that are visible to attackers.

Knight et al., in a report of the DARPA project Self-Regenerative System (SRS) \cite{knight07}, summarize the main features of the Genesis Diversity Toolkit. This tool is one of the most recent approaches that integrates multiple forms of artificial diversity. The goal of the project was to generate 100 diverse versions of a program that were functionally equivalent but for which a maximum of 33 versions had the same deficiency. The tool supports the injection of 5 forms of diversity: Address Space Randomization (ASR), Stack Space Randomization (SSR), Simple Execution Randomization (SER), Strong Instruction Set Randomization (SISR), Calling Sequence Diversity (CSD).

The GENESIS project, also coordinated by Knight's group, explored a complete program compilation chain that applies diversity transformations at different steps to break the monoculture \cite{Williams09}. Diversity transformations are applied compile time, link time, load time,
and runtime. The latter step is the main innovation of GENESIS and relies on the Strata virtual machine technology, which supports the injection of runtime software diversity. This application-level virtual machine realizes two forms of diversification: calling sequence diversity and instruction set diversity.

Jacob et al. \cite{jacob08} propose superdiversification as a technique that integrates several forms of diversification to synthesize individualized versions of programs. The approach, inspired by compilation superoptimization, consists in selecting sequences of bytecode and in synthesizing new sequences that are functionally equivalent. Given the very large number of potential candidate sequences, the authors discuss several strategies to reduce the search space, including learning occurrence frequencies of certain sequences.

Franz \cite{franz10} advocates for massive-scale diversity as a new paradigm for software security. The idea is that today some programs are distributed several million times and all these software clones run on millions of machines in the world. The essential issue is that, even if takes a long time to an attacker to discover a way to exploit a vulnerability, this time is worth spending since the exploit can be reused to attack millions of machines. Franz envisions a new context in which, each time a binary program is shipped, it is automatically diversified and individualized, to prevent large-scale reuse of exploits. The approach relies on four paradigm shifts as enablers for his vision: online software distribution, ultra reliable compilers, cloud computing and good enough performance.

In 2010, Moving Target Defense (MTD) was announced as one of the three ``game-changing'' themes to cyber security the President's Cyber Policy Review announced. The software component of MTD integrates spatial and temporal software diversity, in order to ``limit the exposure of vulnerabilities and opportunities for attack'' \cite{jajodia11}. With such a statement, future solutions for MTD will heavily rely on the integration of various software diversity mechanisms to achieve their objectives.

Inspired by the work of Cohen, who suggested multiple kinds of program transformations to diversify software \cite{cohen93}, Collberg et al. \cite{collberg12} compose multiple forms of diversity and code replacement in a distributed system in order to protect it from remote man-at-the-end attacks. The diversification transformations used in this work are adapted from obfuscation techniques: flatten the control flow, merge or split functions, non-functional code addition, parameter reordering and variable encoding. These transformations for spatial diversity are combined with temporal diversity (when and how frequently diversity is injected), which rely on a diversity scheduler that regularly produces new variants.

Allier et al. recently  proposed to use software diversification in multiple components of web applications \cite{allier14}. They combine different software diversification strategies, from the deployment of different vendor solutions, to fine-grained code transformations, in order to provide different forms of protection. Their form of multi-tier software diversity is a kind of integrated diversity in application-level code.

\subsubsection{Controllers of Automated Diversity}
\label{sec:controller-diversity}

If mixed together and put at a certain scale of automation and size, all kinds of automated diversity need to be controlled. Popov et al \cite{popov2012empirical} provide an in-depth analysis of diversity controllers, showing that diversity controlled with specific diversity management decisions is better than naive diversity maximization.  On the engineering side, several researchers have discussed how to manage the diverse variants of the same program. 

Cox et al. \cite{cox06} introduce the idea of N-variant systems, which consists in automatically generating variants of a given program and then running them in parallel in order to detect security issues. This is different from N-version programming because the variants are generated automatically and not written manually.
The approach is integrated because it synthesizes variants using two different techniques: address space partitioning and instruction set tagging. Both techniques are complementary, since address space partitioning protects against attacks that rely on absolute memory addresses, while instruction set tagging is effective against the injection of malicious instructions. In subsequent work, the same group proposed another transformation that aims at thwarting user ID corruption attacks \cite{nguyen08}.

Salamat and colleagues find a nice name for this concept: ``multi-variant execution environment''~\cite{salamat2008reverse,jackson2011compiler}.
A multi-variant execution environment provides support for running multiple diverse versions of the same program in parallel. The diverse versions are automatically synthesized at compile-time, with reverse stack execution \cite{salamat09,salamat11}.
The execution differences allow some kind of analysis and reasoning on the program behavior. For instance, in \cite{salamat2008reverse}, multi-variant execution enables the authors to detect malicious code trying to manipulate the stack. 

Locasto and colleagues \cite{SidiroglouLK06} introduced the idea of collaborative application communities. 
The same application (e.g. a web server) is run on different nodes. In presence of bugs (invalid memory accesses), each node tries a different runtime fix alternative. If the fix proves to be successful, a controller shared it among other nodes. This healing process contains both a diversification phase (at the level of nodes) and a convergence phase (at the level of the community).

\subsubsection{Summary}
\label{sec:summary-integrated}

Each form of software diversification targets a specific goal (e.g., against a specific attack vector). Many recent work have thus experimented with the integration of multiple forms of diversity in a system, ot benefit from several forms of protection. We have discussed these works here, as well as the specific kinds of controllers that are required to integrate various diversification techniques.

\subsection{Summary}
This section has presented a broad range of contributions on automated software diversity.
They come from different research communities, some of them do not even  use the word diversity. 
However, they all share the same idea that programs and program executions need not be identical. With respect to the rest of this paper, they are fully automated, which is different from the natural diversity discussed in section \ref{sec:managed-natural} and \ref{sec:natural-study} and the managed, yet mostly manual diversity presented in section \ref{sec:managed}.

\section{Diversity as Study Subject}
\label{sec:study-subject}

In this section, we present different works that focus on analyzing and quantifying software diversity and its effects on different aspects of reliability (e.g., fault-tolerance or intrusion-avoidance). 
Contrary to the previous sections, the work presented here is not primarily an engineering contribution, it is not a new technique to support, encourage, or create a new kind of software diversity. 
These approaches all have in common that they consider software diversity as their research subject per se. They simply aim at understanding the deep nature of software diversity from the causes to the implications.
  
First, section \ref{sec:study-n-version} discusses the theoretical models of design diversity and its effects on fault-tolerance. Then, section \ref{sec:natural-study} presents the literature on the analysis of the natural diversity that is found in off-the-shelf components and source code. 

\subsection{Theoretical Modeling Of Design Diversity}
\label{sec:study-n-version}

Failure independence is a critical assumption of the design diversity principle for fault-tolerant critical systems. After the introduction of N-version programming and recovery blocks in the late 70's, a large number of studies have investigated their theoretical foundations and the validity of their assumptions. We discuss the most important studies here.

Design diversity (N-version programming, recovery blocks) was one  of the earliest proposal to leverage diversity and redundancy in software for sake of fault-tolerance. Fault-tolerance is ensured under one essential assumption: the independence of failures among the diverse solutions. Because of the critical impact of this assumption, a large number of papers have investigated the validity of this assumption. While section \ref{sec:design} focused on the principles of design diversity, here we focus on the studies that have evaluated the impact of this approach through empirical studies and  statistical modeling.

Knight and Levenson \cite{knight86} provided the first large-scale experiment that aimed at validating the independence assumption in N-version programming. They asked students to write a program from a single requirements document (for a simple antimissile system) and obtained 27 programs. Each program was tested against 1 million random test cases. The quality of the programs was very high (very few faults), but still there were errors that were found in more than one version (the same error in independently developed programs). A statistical analysis of the results revealed a significant lack of independence between certain errors in the multiple versions of this program. Consequently, the paper was the first major criticism of the effectiveness of design diversity.

Bishop et al.  \cite{bishop86} summarized the results of the PODS project, which aimed at evaluating N-version design on the reliability of software. Their experimental setup is based on the development of three versions of a controller for over-power protection. The requirements document is the same for the three teams, but then they use different methods and languages for the implementation. They concluded that running the three versions, with a voting mechanism, produces a system that is more reliable than the most reliable version and also that back-to-back testing on all three versions is an effective solution to find residual bugs. 

Several pieces of work proposed theoretical frameworks to analyze and quantify the effects of N-version design on reliability.
Eckhardt and Lee \cite{eckhardt85} have developed a theoretical statistical model for evaluating the impact of diversity on fault-tolerance. This model quantifies the effect of joint occurrences of errors on the reliability of the global system. Then, they use this model to explore the conditions under which N-version design can improve fault-tolerance and what are the limits of coincidental errors on the effect of N-version design. Littlewood and colleagues have refined the work of Eckhardt, first by considering the diversity of development methods  \cite{littlewood89}, and more recently by adding further hypotheses and studying two-channel systems \cite{littlewood2012reasoning}. They show that methodological diversity, analyzed as the diversity of development decisions, is very likely to produce behavioral diversity. Popov and Strigini \cite{popov01} proposed another model to analyze the effects of design diversity, in which they rely on data that are more related to physical attributes than previous proposals, making the model more actionable for reliability analysis and prediction. Mitra et al. \cite{mitra99} defined metrics to quantify diversity in N-version designs and highlighted new results about the effectiveness of N versions on software reliability: diversity increases fault tolerance in the presence of common mode failures, as well as self-testing capacities, but the effects of diversity decrease over time. Nicola and Goyal \cite{nicola90} proposed a statistical model that captures the distribution of correlated failures in multiple versions, as well as a combinatorial formula  to predict the reliability of a system running N versions. They analyze the effectiveness of N-version design and demonstrate the need for loose correlations between failures in the N versions.
Hatton \cite{hatton97} evaluates N-version design slightly differently: he proposes a theoretical model to compare the development of a single highly reliable version of a software component, vs. the development of N versions of the component. He concludes that N-version design is good, especially considering our inability to make a really good version.

Kanoun focuses \cite{kanoun99} on a cost analysis of developing 2 diverse versions of the same program. She aims at providing feedback about the overhead of developing the second version, considering one version as the reference. She focuses on working hours records for cost estimates. She observes between 25\% and 134\% overhead depending on the development phase (the highest overhead is for the coding and unit tests, while the lowest if for functional specification). These results confirm other observations from controlled experiments, with actual data from industrial software development.

Partridge and Krzanowski \cite{partridge97} start from the framework of Littlewood and Miller and extend it: they look at the impact of multiple versions beyond failure diversity, including other targets for diversity, such as specializing the performance of some versions for specific tasks.
They evaluate the possibility of an optimal diversity level for reliable software. Partridge and Krzanowski provide an initial attempt to understand the role of software diversity at multiple levels and to systematically quantify diversity in complex systems.

More recently, van der Meulen and Revilla  \cite{Meulen08} analyze the impact of design diversity with thousands of programs that all implement the same set of requirements. 
Those programs come from the UVa Online Judge Website, which proposes a set of programming challenges that can be automatically corrected. Hence, the programs were written by thousands of anonymous programmers attracted by the website concept.
van der Meulen and Revilla use the frameworks of Eckhardt and Lee \cite{eckhardt85} and Littlewood and Miller \cite{littlewood89}. The authors classify different categories of faults that occur in different versions, and then, through random selections of pairs of versions, evaluate the reliability of the system (assuming that the system does not fail if one of the versions does not fail). They confirm that N-version design is more effective when different versions fail independently and that the diversity of programming language has a positive effect (programmers make different faults and different kinds of faults, with different languages). 
Given the size of their dataset, the authors really stress the statistical validity of their findings.

Salako et al. \cite{salako13} question the independent sampling assumption posed by the models of Eckhardt and Lee \cite{eckhardt85} and Littlewood and Miller \cite{littlewood89}. They analyze the consequences of violating this assumption and evaluate the opportunity of using different versions of a program (not developed independently) to build fault-tolerant systems. Their results confirm the important influence of independence on diversity. Yet, they also open the discussion about different forms of independence and different processes that can be applied to mitigate the influences between different versions.

A large number of theoretical and empirical studies have dissected the foundations of design diversity. We have summarized these works here and discussed how they have contributed to a finer grain understanding of the conditions for effective design diversity. 

\subsection{Study of Natural Software Diversity}
\label{sec:natural-study}

``Natural software diversity''  is any form of software diversity that spontaneously emerges from software development. The emergence comes from many factors such as the market competition, the diversity of developers, of languages or of execution environments.  
In Section \ref{sec:managed},
 we have discussed how natural diversity can be used to establish reliable software systems (Section \ref{sec:managed-natural}).
In this section, we resume on natural diversity and discuss the literature that studies and describes this existing natural diversity.
The different studies presented in this section explore different kinds of software diversity: in software components, in source code, as well as in the social behaviors in open source communities.

Gashi et al. \cite{gashi04} have studied bug reports for 4 off-the-shelf SQL servers (Oracle 8.0.5, Microsoft SQL, PostgreSQL 7.0.0 and Interbase 6.0), to understand whether these solutions could be good candidates for fault-tolerance, i.e., exhibit failure diversity. The study consisted in selecting bugs for each of the servers, collect the test cases that trigger the bug on a server and run them on the other servers to check whether the other solutions present the same bug. Following this protocol, for a total of 181 bugs, they observed that only 4 were bugs in two versions simultaneously, and no bug was found in more than 2 versions. They emphasize that the diversity of solutions is major asset for forward error recovery, since it is possible to copy the state of a correct database in a failed one. They have proposed to use this natural diversity to design an architecture for a fault-tolerant database management system \cite{gashi07}.

Barman et al. \cite{barman09} focus on host intrusion detection systems (HIDS) deployed on all machines of entreprise networks. The ability of an IDS to detect intrusions depends on different thresholds that should depend on each user, yet these thresholds are usually set to the same value on each machine, because of a lack of guidelines about how to configure them. The authors analyze the impact of this monoculture of HIDS, showing that it provides very poor results in terms of intrusion detection. These poor results are mainly because the behavior of users are so diverse that they HIDS should also have diverse configurations to be effective. Then, the authors experiment with increasing configuration diversity and observe a clear benefit to reduce the number of missed detections.

Koopman and De Vale \cite{koopman99} evaluate the diversity of POSIX operating systems, using a robustness metric based on failure rates. The authors compare 13 implementations of POSIX. They use the Ballista testing tool to generate large quantities of robustness test cases that they run on each version. This reveals between 6\% and 19\% of failure rate. Then, the authors perform a multi-version comparison to analyze the diversity of failures and thus the usability of these POSIX versions for N-version fault-tolerance. The results demonstrate that multi-versions can be used to increase robustness, yet, with the 2 most diverse solutions, there is still a 9.7\% common mode failure exposure for system calls.

Han et al. \cite{han09} analyze the diversity of off-the-shelf components with respect to their diversity of vulnerabilities. They provide a systematic analysis of the ability of multi version systems to prevent exploits. The study is based on 6000 vulnerabilities published in 2007. The main result is that components available for web servers are diverse with respect to their vulnerabilities and cannot be compromised by the same exploit.
Consequently, all these components can run on multiple operating systems in order to increase diversity. They conclude that the natural diversity of off-the-shelf software applications is beneficial to build attack tolerant systems.

Some recent work study the natural diversity or redundancy that emerges in large-scale source code. Gabel and Su \cite{gabel10} analyze uniqueness in source code through the analysis of 6000 programs covering 420 million lines of code. The authors focus on the level of granularity at which diversity emerges in source code. Their main finding is that, for sequences up to 40 tokens, there is a lot of redundancy. Beyond this (of course fuzzy) threshold, the diversity and uniqueness of source code appears. 
Jiang and Su \cite{jiang09} propose an approach for the identification of functionally equivalent source code snippets in large software projects. This approach consists in extracting code snippets of a given length, randomly generating input data for these snippets and identify the snippets that produce the same output values (which are considered functionally equivalent, w.r.t the set of random test inputs). They run their analysis on the Linux kernel 2.6.24 during several days and find a large number of functionally equivalent code fragments, most of which are syntactically different. Both studies explore the tension between redundancy and diversity that exists in software. 

Mendez et al. \cite{mendez13} analyze the diversity in source code at the level of usages of Java classes. They analyze hundreds of thousands of Java classes, looking for type usages, i.e. sets of methods called on an object of a given type. They find 748 classes with more than 100 different usages of the API, the most extreme case being the \texttt{String} of the Java library, for which they found 2460 different usages. This reveals a very high degree of usage diversity in object-oriented software.

Diversity also emerges in social behaviors in open source software development. In this area, Posnett et al.  \cite{posnett13}  analyze the focus of developers (whether they contribute to few or many artifacts) and the ownership (to what extent an artifact is ``owned'' by one or several developers). Through an analogy with predator-prey relations, they set up entropy measures to quantify the diversity in focus and ownership. They observe high levels of diversity in open source projects, and also demonstrate that these entropy metrics have good predictive properties: focused developers introduce less defects, while artifacts that receive contributions from several developers tend to have more defects. 
Vasilescu et al. \cite{vasilescu13} studied the development of the GNOME community and observed diversity both from the point of view of contributors (how diverse are the activities of different project contributors) as well as from the point of view of project (how diverse are the activities going on in different GNOME projects).

Software diversity spontaneously emerges through multiple phenomena. In this section we have discussed the methods to study  these different phenomena, as well as the experimental procedures that have been implemented to analyze the impact of this specific form of software diversity. These recent studies illustrate how the analysis of complex diversification processes must leverage techniuqes from multiple domains ranging  from software analysis, data mining, statistics to threat models and exploit replication.

\subsection{Summary}
This section has presented two main areas in the analysis and the theoretical modeling of software diversity and its impact. The first part provided an overview of 3 decades of works that analyzed N-version programming and proposed several statistical methods and foundational assumptions that underly the effectiveness of this technique for fault-tolerant software systems.
The second part discusses novel work that analyze the implication and the effectiveness of natural software diversity (as presented in section \ref{sec:managed-natural}) for building resilient systems.

\section{Conclusion}

In this paper, we provided a global picture of the software diversity landscape. We decided to  broaden the standard scope of diversity, in order to give a very inclusive vision of the field and, hopefully, a better understanding of the nature of software diversity. The survey gathered work from various scientific communities (security, software engineering, programming languages), which we organized around one dimension: the diversity engineering technique (managed, automated, natural).

Looking at all these works from a temporal perspective, we realize that the interest for diversity has always existed in the last 40 years. The latest studies even discover phenomena of natural diversity emergence, i.e. diversity is observed but the processes that led to its presence are unknown. We believe that harnessing this natural diversity will be an essential step in the future of software diversification. This could be the intermediate step towards the amplification of natural diversity. Indeed, diversity in natural complex systems is never explicitly developed, but emerges as a side effect of other phenomena. For example, biodiversity at different scales of ecosystems, emerges as the result of sexual reproduction, mutation, dispersal and frequency-dependent selection \cite{de2009,melian2010}. To this extent, the main area of future work is to identify the software engineering principles and evolution rules that drive the emergence and the constant renewal of diversity in software systems. In other words, can we engineer open-ended software diversification?

\section*{Acknowledgements}
We would like to thank Paul Amman, Benoit Gauzens and Sebastian Banescu for their valuable feedback on this paper. This work is partially supported by the EU FP7-ICT-2011-9 600654 DIVERSIFY project.

\bibliographystyle{ACM-Reference-Format-Journals} 
\bibliography{u4f673723ea638}

\end{document}